\documentclass[conference,final]{./IEEEtran}
\usepackage{amsfonts} % if you want blackboard bold symbols e.g. for real numbers
\usepackage{graphicx} % if you want to include jpeg or pdf pictures and put it in a flexible way
\usepackage{amssymb}% http://ctan.org/pkg/amssymb

\usepackage{multirow}% multirow in a tabular
\usepackage{eso-pic}% http://ctan.org/pkg/eso-pic
\usepackage[ruled,vlined]{algorithm2e}
\usepackage{cite}
\usepackage[framemethod=TikZ]{mdframed}
\usepackage{xcolor}

\usepackage{amsmath, cases}
\usepackage[bookmarks=false,draft]{hyperref} % IEEE final doesn't allow bookmarks
\usepackage[english]{babel}
\addto\captionsenglish{\renewcommand{\figurename}{Fig.}}
\addto\captionsenglish{\renewcommand{\tablename}{TABLE}}
\addto\extrasenglish{ % capitalize the label, like section
    
}
\addto\extrasenglish{ % capitalize the label, like section
    
}

\def\equationautorefname#1#2\null{%fix autoref no equation brackets
  Equation#1(#2\null)%
}

\usepackage{tabularx,booktabs}
\newcolumntype{C}[1]{>{\Centering}m{#1}}

\usepackage[font=small]{caption}

\usepackage{tikz}% draw network protocol
\usepackage{etoolbox}

\newcommand{\circled}[2][]{% use circled number
  \tikz[baseline=(char.base)]{%
    \node[anchor=text, shape=circle,draw, inner sep=0pt, minimum size=0.5em] (char){#1\strut};
    \node at (char.center) {\makebox[0pt][c]{#2}};}}
\robustify{\circled}
\newcommand{\epc}{\textit{EPC Gen2}\xspace}
\newcommand{\readme}{\textit{ReaDmE}\xspace}

\begin{document}

\title{\readme: Read-Rate Based Dynamic Execution Scheduling for Intermittent RF-Powered Devices}

\author{\IEEEauthorblockN{Yang Su}
\IEEEauthorblockA{\textit{School of Computer Science} \\
\textit{The University of Adelaide}\\
Adelaide, Australia \\
yang.su01@adelaide.edu.au}
\and
\IEEEauthorblockN{Damith C Ranasinghe}
\IEEEauthorblockA{\textit{School of Computer Science} \\
\textit{The University of Adelaide}\\
Adelaide, Australia \\
damith.ranasinghe@adelaide.edu.au}
}

\maketitle
\footnotetext{© 2021 IEEE.  Personal use of this material is permitted.  Permission from IEEE must be obtained for all other uses, in any current or future media, including reprinting/republishing this material for advertising or promotional purposes, creating new collective works, for resale or redistribution to servers or lists, or reuse of any copyrighted component of this work in other works.}
\begin{abstract}
This paper presents a method for remotely and dynamically determining the execution schedule of long-running tasks on intermittently powered devices such as computational RFID. Our objective is to prevent brown-out events caused by sudden power-loss due to the intermittent nature of the powering channel. We formulate, validate and demonstrate that the read-rate measured from an RFID reader (number of successful interrogations per second) can provide an adequate means of estimating the powering channel condition for passively powered CRFID devices. This method is attractive because it can be implemented without imposing an added burden on the device or requiring additional hardware. We further propose \readme, a dynamic execution scheduling scheme to mitigate brownout events to support long-run execution of complex tasks, such as cryptographic algorithms, on CRFID. Experimental results demonstrate that the \readme method can improve CRFID's long-run execution success rate by 20\% at the critical operational range or reduce time overhead by up to 23\% compared to previous execution scheduling methods.
\end{abstract}

\section{Introduction}
Rapid evolution of microelectronic technologies is still ongoing, more and more components and functions are being integrated into tiny chips with surprisingly low power consumption. High integration and relatively low energy requirements make it possible for electronic systems to run purely on harvested energy, such as computational RFID (CRFID) which operate on harvested radio wave energy. In addition to the identity authentication and simple data storage defined in the \epc standard\cite{epcglobal2015inc}, CRFID devices may also perform general-purpose computations \cite{wickramasinghe2014windware,RanasinghePlosOneWearableSensor}, inertial sensor sampling\cite{camera2020experimental,6347459}, cold supply chain monitoring for food and vaccine transportation \cite{monteleone2017novel}, brain-computer/machine interfacing \cite{dementyev2013wearable}, medication monitoring \cite{payne2020medication}, aircraft maintenance \cite{aantjes2017fast}, transmitting speech \cite{talla2017battery} and streaming video \cite{saffari2019battery} to name a few. CRFID devices send back results via the RFID backscatter communication channel~\cite{thomas2020backscatter} and, more importantly, are completely powered by the harvested energy.

\begin{figure}[!t]
    \centering
    \includegraphics[width=\linewidth]{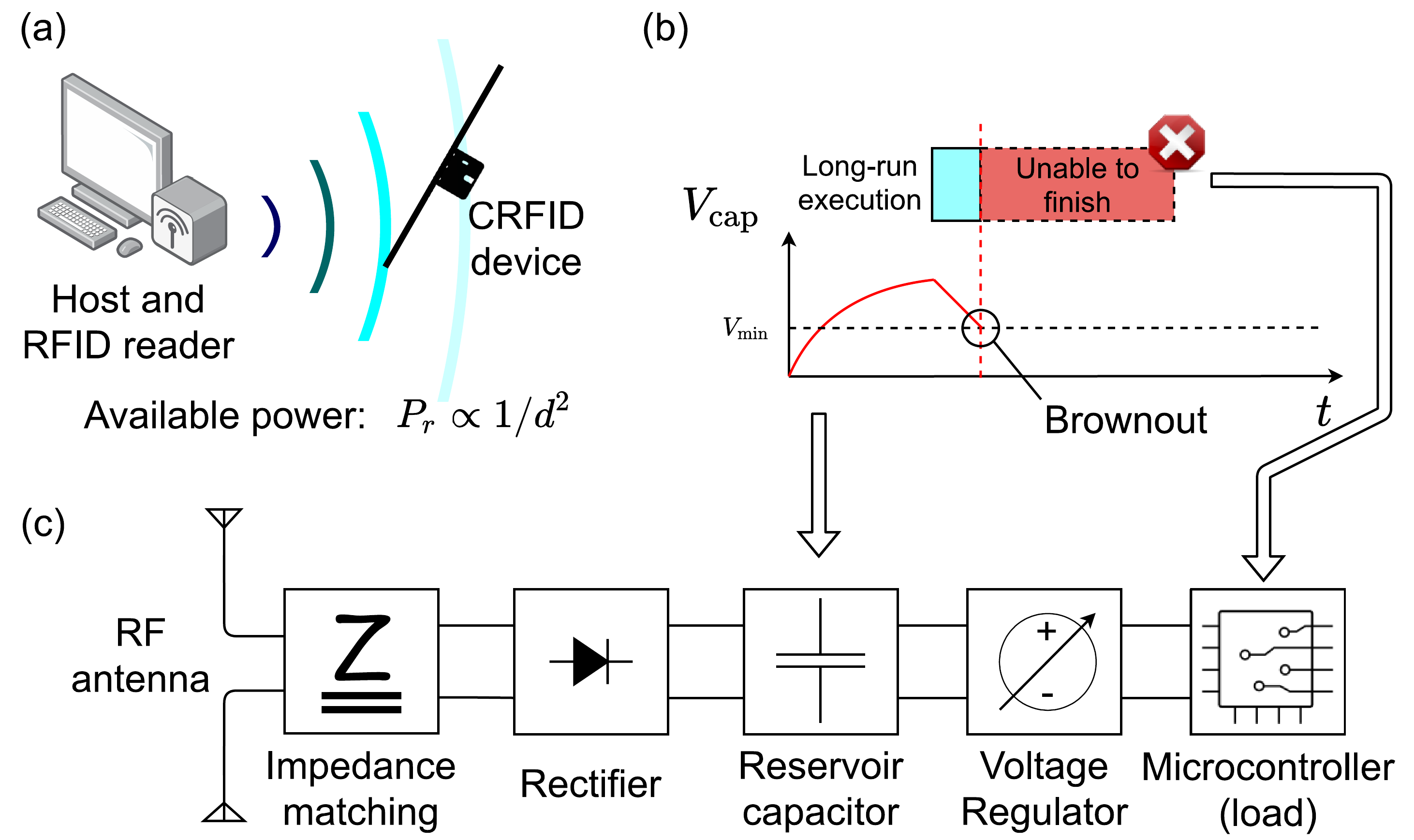}
    \caption{(a) Due to the properties of  RF energy propagation, at a certain distance away from the reader (b) CRFID devices may be unable to finish the long-run execution of a task due to a brownout event. Here, the voltage across the reservoir capacitor $V_{\rm cap}$ drops below the minimal operational voltage of the Mcirocontroller $V_{\rm min}$. (c)  Shows the architecture of a typical passively powered device operating on harvested RF energy.} 
    \label{fig:ReaDmE_F1}
\end{figure}

However, the power that a CRFID device receives from a radiating RFID reader antenna reduces as distance increases\cite{sun2017fully,friis1946free,nikitin2010phase}, as illustrated in~\autoref{fig:ReaDmE_F1}~(a), due to path loss, shadowing, scattering and radio wave absorbing effects from obstacles \cite{dong2015novel}. Further, power harvesting devices typically use burst-charge cycles based on a charge pump to deliver packets of energy to power the devices. These bursts become intermittent at larger operating distances. Consequently, these devices operate over a sequence of short-term bursts and work in so-called intermittent powering conditions. Therefore, CRFID type devices may not always be able to support long-run execution of tasks, as shown in \autoref{fig:ReaDmE_F1}~(b). 

Recent studies have considered methods for operating under intermittent powering conditions to support the long-run execution of tasks, such as cryptographic algorithms. However, existing methods normally require specific hardware support \cite{sample2008design, buettner2011dewdrop} or introduce additional execution overhead \cite{su2018secucode,salajegheh2009cccp}. In this paper, we investigate the following question: 

\vspace{1mm}
\begin{mdframed}[backgroundcolor=black!4,rightline=false,leftline=false,topline=false,bottomline=false,roundcorner=2mm]
	How can CRFID devices make efficient use of harvested power with minimal cost to the device, in terms of hardware requirements and implementation overhead, to  achieve long run execution of tasks? 
\end{mdframed}
\vspace{1mm}

\begin{table*}[hbt!]
\centering
\caption{Comparison of Related Works.}\label{tab:related_work}
\resizebox{\textwidth}{!}{
    \begin{tabular}{lllll} 
    \toprule
    Study & Scheduling method & Power measurement & Scheduling action & Hardware requirements  \\ 
    \hline\hline
    WISP 4.1 (HwFixed) \cite{sample2008design} & Programmatic  & On-device voltage & Pre-task decision & On-device voltage supervisor  \\
    Dewdrop\cite{buettner2011dewdrop}        & Dynamic & On-device voltage &  In-task decision & On-device ADC and passive voltage multiplier      \\
    IEM \cite{su2018secucode} & Programmatic  & None & In-task sleep & None\\
    Check-pointing\cite{maeng2017alpaca,ransford2011mementos,hicks2017clank,salajegheh2009cccp} & None & None & In-task check-pointing & On-device NVM or reader coordination\\
    ReaDmE (\textbf{\textit{this paper}}) & Dynamic & Reader-driven & In-task sleep & None\\
    \bottomrule
    \end{tabular}
}
\end{table*}

Our main contributions from our efforts to address the above research question are as follows: 
\begin{itemize}
    \item We formally analysed the relationship between the number of successful interrogating rounds (\textit{read-rate}) and the time required for a CRFID device to harvest adequate energy (charge time) and subsequently formulated an analytical expression linking these two quantities.
    \item To the best of our knowledge, this work is the first to attempt to achieve a reader-driven \textit{dynamic} execution scheduling method for intermittent RF powered devices utilising the simplicity of read-rate measurements and with minimal implementation cost to the device.
    \item We designed, implemented and evaluated a reader-driven read-rate based dynamic execution scheduling scheme for long-run executions suited for intermittent RF powered devices.
\end{itemize}

The rest of this paper is organised as follows:
We first review related works in \autoref{sec:related-work}, then provide the necessary background knowledge to understand an intermittently powered system in \autoref{sec:background}. We discuss the principle of our proposed method in \autoref{sec:principle}, followed by the detail design of the \readme scheme. We implement and evaluate \readme in Sections \ref{sec:implementation} and \ref{sec:experiment}, and conclude our work and discuss future research directions in \autoref{sec:conclusion}.

\section{Related work}\label{sec:related-work}
The intermittent powering condition for CRFID is first considered in the original WISP design. In \cite{sample2008design}, Sample {\it et al.} identified that, the harvested energy may not always be sufficient to accomplish the desired computational task. To overcome this issue, a dedicated hardware voltage supervisor is employed to monitor the harvested energy level. The voltage supervisor is a comparator with an internal voltage reference $V_{\rm ref}$. The device's microcontroller unit (MCU) may poll the voltage supervisor before running any power-intensive tasks, such as complex computations or sensor sampling. If the harvested voltage level is below the threshold, the MCU puts itself into a deep sleep state to accumulate energy. Once the voltage level reaches $V_{\rm ref}$, the voltage supervisor sends an interrupt to wake-up the MCU. The CRFID device can then safely launch the power-intensive task. Since the CRFID wakes up at a fixed voltage, via the hardware voltage supervisor, this method is referred to as \textit{HwFixed}~\cite{sample2008design}. A limitation of this setup is that it relies on the dedicated hardware voltage supervisor available only on WISP 4.1DL. Notably, this supervisor is absent in the more recent version of the WISP\cite{Wisp5}.

Buettner  {\it et al.} proposed the Dewdrop~\cite{buettner2011dewdrop} execution model to \textit{prevent} a brownout event due to  unreliable powering; the method monitors available harvested power and executes tasks only when they are likely to succeed. Dewdrop utilises a dynamic on-device task scheduling method but requires the overhead of collecting samples of the harvester voltage and task scheduling by the device's application code. Further, Dewdrop is only suitable for CRFID devices equipped with a passive charge pump, such as WISP~4.1~\cite{sample2008design} since Dewdrop requires an analog to digital converter (ADC) to directly measure the charging rate of the reservoir capacitor (the charge storage element). In the follow-up, WISP version 5.1, the passive charge pump is replaced with an active charge pump (S-882z) and the reservoir capacitor is only connected to the load when $V_{\rm cap}$ developed across the capacitor exceeds the reference voltage of 2.4~V $V_{\rm ref}$. Consequently, in WISP~5.1, the voltage delivered to the microcontroller is a sharp step-up, rather than a ramp-up function related to harvested power. Therefore, the charging-up process cannot be directly monitored using the technique in the Dewdrop model.  

Su {\it et al.}~\cite{su2018secucode} proposed the intermittent execution model (IEM) as a part of the SecuCode executable code dissemination scheme. IEM is designed to prevent brownout by fragmenting the power-intensive tasks into small sub-tasks and sandwiching them with low power sleep modes (LPM), thus allowing the CRFID devices to recharge regularly. IEM makes use of the MCU's internal timer to wake the system up from LPM. Although SecuCode is designed to perform  firmware updates, securely, the IEM execution schedule is hard-coded as a part of the immutable bootloader. IEM cannot dynamically adjust the schedule once the CRFID device is deployed. Consequently, the fixed-term LPM states could be incurred unnecessarily, with resulting time overheads, despite the availability of a good powering channel.

Checkpointing is another well-studied technique to handle unpredictable intermittent powering conditions. By regularly saving the run-time data into a device's non-volatile memory (NVM)\cite{maeng2017alpaca,ransford2011mementos,hicks2017clank} or uploading to a server after encryption~\cite{salajegheh2009cccp} (\textit{i.e.,} the checkpoint), the system can resume from the last intact checkpoint whenever a power failure occurs. However, writing to NVM is energy intensive \cite{salajegheh2009cccp} and raises security concerns~\cite{krishnan2018exploiting,salajegheh2009cccp}. Off-device checkpointing methods \cite{salajegheh2009cccp} requires coordination between the reader and the device, as well as time and energy overheads (e.g., for data transmissions and encryption functions).

Notably, received signal strength indicator (RSSI) has been investigated in numerous studies \cite{buffi2018rssi,zhang2012blink} for communication channel characterisation, but no studies to date have considered using RSSI for task scheduling.

\vspace{2mm}
\noindent\textit{Summary.~}The existing methods for dealing with power loss are compared to in TABLE~\ref{tab:related_work}. In contrast to current approaches, we propose using remote measurements from the transceiver (RFID reader in case of reading and writing to CRFID devices) to characterise the available power at the CRFID device and allow the transceiver to determine, at run time, the dynamic execution schedule for a device.

\section{Background}\label{sec:background}
In this sections, we briefly describe the background information pertinent to our study.

\subsection{Typical energy-harvesting device architecture}\label{sec:background_architecture}

A generic power supply design for an RF energy harvester architecture is depicted in \autoref{fig:ReaDmE_F1}~(c). The RF antenna is driven by the electromagnetic waves, starting from the left-hand side, and pushes charges into the impedance-matching network. The impedance matching technique maximises energy harvesting efficiency at a specific frequency. The rectifier ensures charges only flow along one direction to produce a direct current (DC). The generated charges are buffered into an energy storage component, normally a reservoir capacitor. The buffered electricity is regulated to the desired voltage before delivering to the load, such as the MCU of a CRFID device.

\subsection{Execution model in HwFixed scheme}\label{sec:background_operation}

\begin{figure}[!ht]
    \centering
    \includegraphics[width=\linewidth]{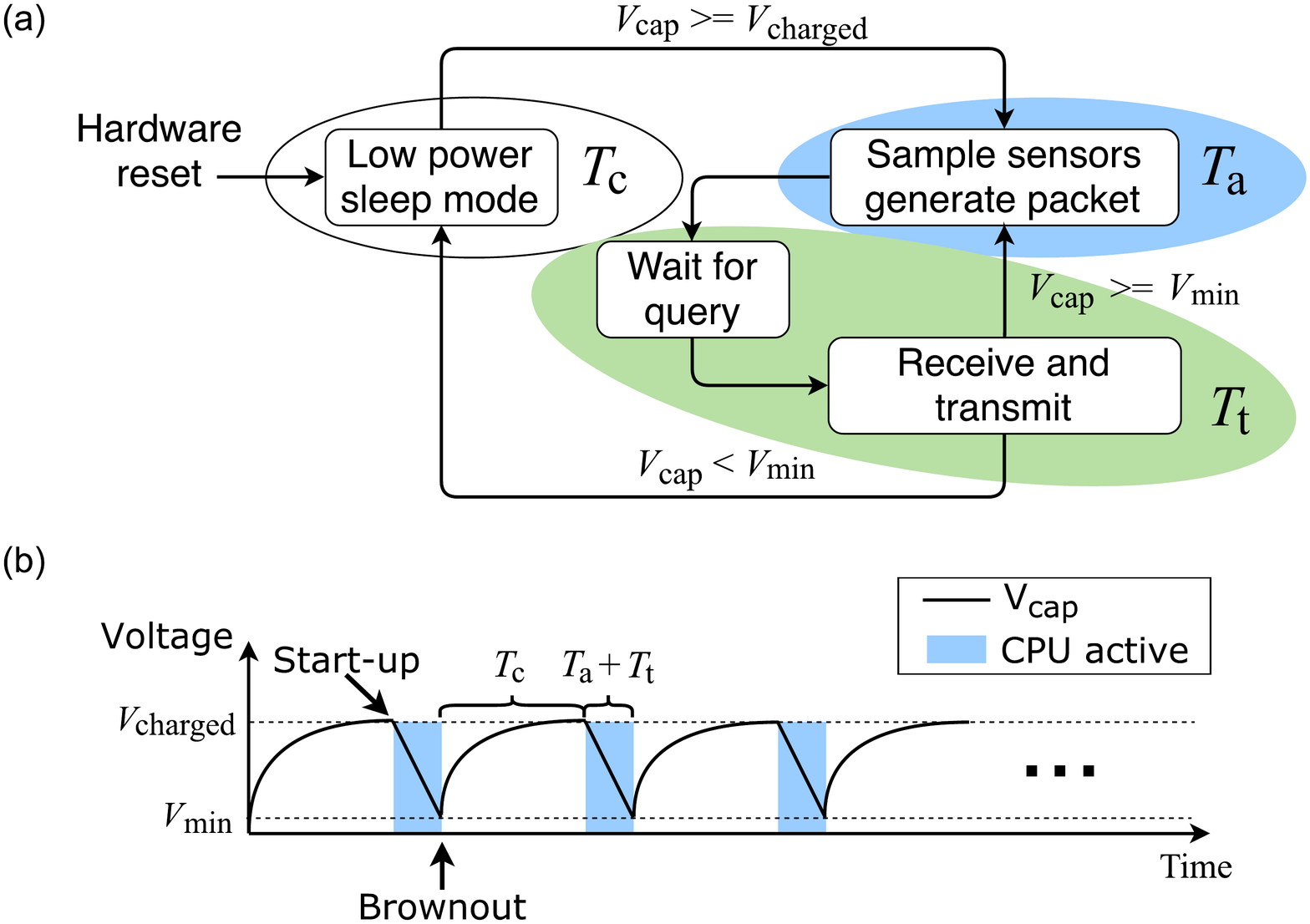}
    \caption{(a) Operational power cycle of a CRFID device and (b) charge burst operational mode.}
    \label{fig:CRFID_Burst}
\end{figure}

Normally, CRFID devices are under intermittent powering conditions as there is no guarantee that the energy harvester can always supply the load with the required power. \autoref{fig:CRFID_Burst} describes the underlying execution model adopted in WISP version 4.1DL and the resulting execution cycles from intermittent powering. On startup, the CRFID device first charges its reservoir capacitor. We denote the time elapsed for the capacitor to be charged as $T_{\rm c}$. A conceptual plot of voltage at the reservoir capacitor $V_{\rm cap}$ versus time is illustrated in \autoref{fig:CRFID_Burst}~(b).  Once the voltage at the reservoir capacitor $V_{\rm cap}$ reaches a threshold $V_{\rm charged}$---see \autoref{fig:CRFID_Burst}~(b)---the MCU is booted up. The MCU may execute code, sample sensors and prepare the up-link packet. This process may take time $T_{\rm a}$ to finish, with `a' denoting that the MCU is in active mode. Subsequently, the CRFID enters the `Wait for query' state until the reader instructs it to backscatter the data packet generated earlier. The waiting and backscattering fall within the time frame $T_{\rm t}$.

 Notably, $V_{\rm cap}$ is periodically charged and discharged. If nothing unexpected happens, the period of one charge-discharge cycle, the intermittent power cycle (IPC) following the definition in \cite{su2018secucode}, is dominated by the three previously mentioned time variables: $T_{\rm c}$, $T_{\rm a}$ and $T_{\rm t}$.

\section{Read-Rate Based Dynamic Execution Scheduling Method}\label{sec:principle}%powering condition index}
We recognise that it is possible for a Server that controls the RFID reader in our setting, to estimate the time to harvest adequate energy at a CRFID device and hence, determine the length of the IPC dynamically. Therefore, in contrast to an apriori chosen intermittent operating setting, as in~\cite{su2018secucode}, we propose the dynamic selection of IPC settings by the Server to reduce unnecessary delays in task execution. 

\subsection{Theory and proof} \label{sec:thery_n_proof}
\noindent\textbf{Observation.~}The number of successful CRFID device interrogations per second \cite{bothe2016improving}, commonly known as the the read-rate $R_{\rm read}$, reduces as the interrogation range $d$ increases.

\vspace{2mm}
\noindent\textbf{Proposition.~}$R_{\rm read}$ can be employed to infer information related to available harvested power by a CRFID device at a given distance. At increasing distances, where harvesting RF power is increasingly difficult, the $R_{\rm read}$ is largely determined by the IPC. We assume the time spent executing code and communicating over the  RFID channel is invariant at different powering conditions. Therefore, $R_{\rm read}$ will be dominated by the charging time $T_{\rm c}$.

Therefore, we expect a CRFID devices charging time can be expressed as a function $f$ of its read-rate: $T_{\rm c} = f(R_{\rm read})$.

\vspace{2mm}
\noindent\textbf{Proof.~}Consider an ideal radio wave propagation environment. The relationship between the transmitted power $P_t$ by an RFID reader antenna and the received power $P_r$ by a CRFID device is expressed by the Friis transmission equation \cite{friis1946free}, illustrated in equation \autoref{eqn:Friis}. The available power at the receiver $P_r$ is determined by transmitted power $P_t$, transmitter antenna gain $G_t$, receiver antenna's gain $G_r$, the radio frequency wavelength $\lambda$ and the distance from the transmitter to the receiver $d$.

\begin{equation}
    \label{eqn:Friis}
    P_r = \frac{P_t G_t G_r \lambda^2}{(4 \pi d)^2}
\end{equation}

We can see that the received power $P_r$ is inversely proportional to the square of distance $d$, assuming all other parameters are constant.

Considering losses at the RF front-end and  burst-charge pump, $P_{\rm charge} = \eta P_r$, where $\eta$ is the efficiency factor of the energy harvester. Now we can express the energy stored during a charging cycle $E_{\rm stored}$ as:

\begin{equation}
    \label{eqn:chargeTime}
    E_{\rm stored} = \frac{1}{2} C (V_{\rm charged}^2 - V_{\rm min}^2) = P_{\rm charge} \times T_{\rm c}
\end{equation}

As discussed in \autoref{sec:background_operation}, the length of a charging cycle is $T_{\rm c}$, $C$ is the capacitance of the reservoir capacitor, $V_{\rm charged}$ indicates the voltage at the end of the charging cycle when the MCU starts up, $V_{\rm min}$ is the minimum voltage a CRFID device requires for normal operation. Since $C$, $V_{charged}$ and $V_{min}$ are constants, $T_{\rm c} \propto \frac{1}{P_{r}} \propto d^2$. This explains our observation that a CRFID device takes a longer time charge and respond to interrogation at longer operating distances.

During $T_{\rm c}$, a CRFID device's power consumption is non-zero, although the MCU is in the LPM; further, the reservoir capacitor has leakage power. We introduce a new term, $P_{\rm sleep}$, to represent power losses during $T_{\rm c}$. On the other hand, $P_{\rm charge}$ would still be present when the MCU is active $T_{\rm a}$. Therefore we formulate $T_{\rm c}$  as in \autoref{eqn:Tcharge1}; if the charge power $P_{\rm charge}$ is greater than the system power consumption during LPM $P_{\rm sleep}$, the charge time $T_{\rm c}$ is given by the ratio of the energy required to fully charge the storage component $E_{\rm storage}$ over the difference between the charge power $P_{\rm charge}$ and the system power consumption under LPM $P_{\rm sleep}$. Otherwise, if $P_{\rm charge}$ is lower than the $P_{\rm sleep}$, voltage across the reservoir capacitor $C$ can never build up, and thus $T_{\rm c}$ is infinite. The time $T_{\rm a}$ that the MCU (and thus the CRFID device) is active is given by \autoref{eqn:Tactive1}; the ratio of energy stored over the difference between the system power consumption of active mode $P_{\rm active}$ and the charge power $P_{\rm charge}$, if $P_{\rm charge}$ is below $P_{\rm active}$. Otherwise, $T_{\rm a}$ is the time elapsed to execute the CRFID's application code (case in \autoref{eqn:Tactive2}).

\begin{numcases}{T_{\rm c} = }
    \dfrac{E_{\rm stored}}{P_{\rm charge}-P_{\rm sleep}} &, $\text{if } P_{\rm charge} > P_{\rm sleep}$ \label{eqn:Tcharge1}\\
    \infty &, $\text{if } P_{\rm charge} \leq P_{\rm sleep}\label{eqn:Tcharge2}$
\end{numcases}

\begin{numcases}{ T_{\rm a} = }
    \dfrac{E_{\rm stored}}{P_{\rm active}-P_{\rm charge}} &, $\text{if } P_{\rm active} > P_{\rm charge}$ \label{eqn:Tactive1}\\
    T_{\rm execute} &, $\text{if } P_{\rm active} \leq P_{\rm charge}$ \label{eqn:Tactive2}
\end{numcases}

As expressed in \autoref{eqn:RR1}, the $R_{\rm read}$ is 0 if $P_{\rm charge}<P_{\rm sleep}$ as the reservoir capacitor takes an infinite time to be charged. Consequently, the CRFID never responds to an RFID reader's query. Alternatively, the CRFID device may respond at $R_{\rm read}$ = $\frac{1}{T_{\rm c}+T_{\rm a}+T_{\rm t}}$. That is one closed loop consisting of charge, activation and transmission as illustrated in \autoref{fig:CRFID_Burst}~(c).
For a successful RFID inventory, $T_{\rm a}$ and $T_{\rm t}$ will be constant; the only variable is the charging time $T_{c}$.

\begin{numcases}{R_{\rm read}=}
    0 &, $\text{if }P_{\rm charge}<P_{\rm sleep}$ \label{eqn:RR1}\\
    \dfrac{1}{T_{\rm c}+T_{\rm a}+T_{\rm t}} &, $\text{if }P_{\rm charge} \geq P_{\rm sleep}$ \label{eqn:RR2}
\end{numcases}

We can easily express $T_{\rm c}$ as a function of $R_{\rm read}$, shown in \autoref{eqn:tc_fn_of_RR},  while assuming $K = T_{\rm a} + T_{\rm t}$ is a constant.

\begin{equation}
    T_{\rm c} = {(R_{\rm read})}^{-1} - K \label{eqn:tc_fn_of_RR}
\end{equation}

Hence, we can express the charging time $T_{\rm c}$  as a function of read-rate $R_{\rm read}.~\blacksquare$

\subsection{\readme design}\label{sec:design}
\begin{figure}[!ht]
    \centering
    \includegraphics[width=.9\linewidth]{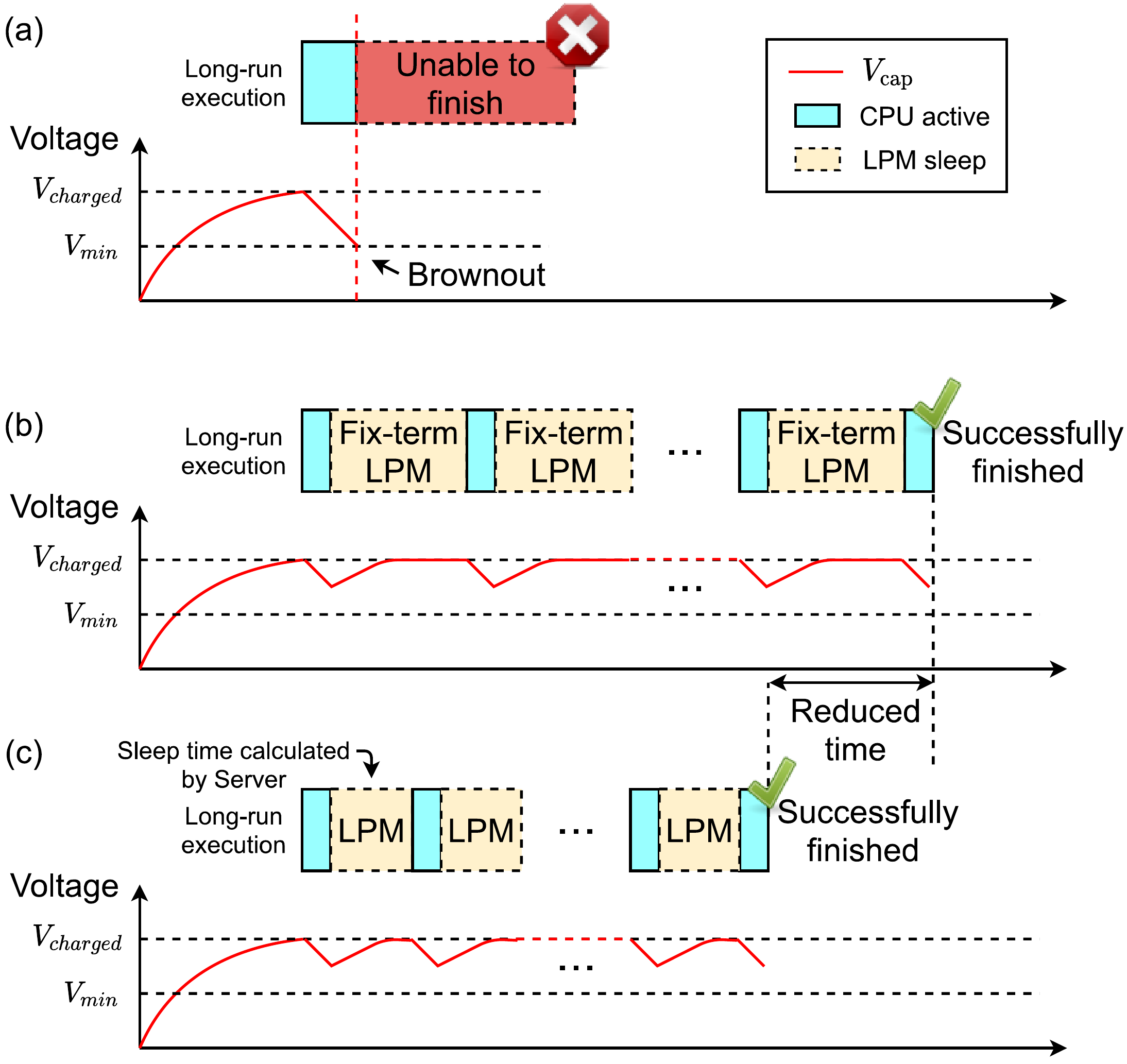}
    \caption{Distinguishing (a) CEM (continuous execution model), (b) IEM (intermittent execution model) and (c) \readme.}
    \label{fig:ReadRate_IEM_compare}
\end{figure}

We develop our \readme method by building upon the IEM scheme in \cite{su2018secucode} since it is the only candidate requiring no specific hardware modifications and is realised with the popular CRFID device, WISP 5.1 LRG\cite{Wisp5}. IEM is  part of the SecuCode bootloader as discussed in \autoref{sec:related-work}. Our proposed \readme is distinguished from the continuous operation model (CEM) and IEM in \autoref{tab:related_work} and  \autoref{fig:ReadRate_IEM_compare}. 

As shown in \autoref{fig:ReadRate_IEM_compare}~(a), the CEM tries to finish a long-run execution in a single burst, but it fails due to the brownout, where the reservoir capacitor voltage $V_{\rm cap}$ cannot hold the system's minimal voltage requirement. 

The IEM (\autoref{fig:ReadRate_IEM_compare}~(b)) fragments the long-run execution into multiple small sub-tasks and sandwiches them with LPM sleep with a fixed duration. Therefore $V_{\rm cap}$ can be maintained above the $V_{\rm min}$ until the long-run execution successfully finishes. However, since the duration of LPM is fixed, this may incur unnecessary time overhead. 

Our proposed \readme (\autoref{fig:ReadRate_IEM_compare}~(c)) dynamically adjusts the duration of LPM sleep according to the $R_{\rm read}$ reported by the RFID reader. Also, if the powering condition is extremely poor \readme could assign longer LPM sleep time to allow for CRFID recharge.

\section{\readme implementation}\label{sec:implementation}
\begin{figure}[!ht]
    \centering
    \includegraphics[width=.7\linewidth]{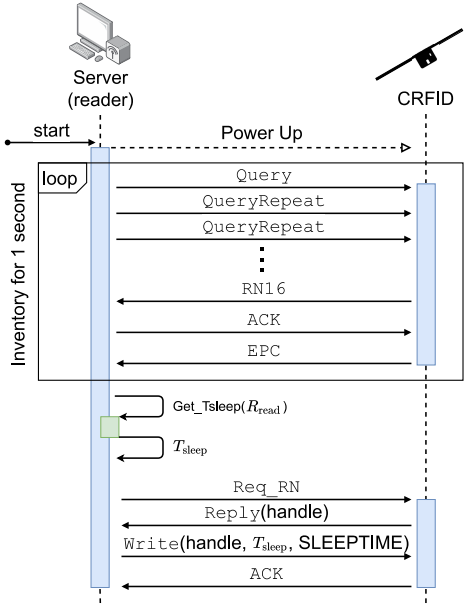}
    \caption{The implementation of \readme over the \epc protocol.}
    \label{fig:ReaDmE_implementation}
\end{figure}
As shown in \autoref{fig:ReaDmE_implementation}, we visualise the implementation of \readme over the \epc protocol in a sequence diagram. We can identify three stages: 

\vspace{2mm}
\noindent\textbf{Stage 1.~}The reader powers up the CRFID device, and counts the $R_{\rm read}$ over a fixed period, for example, one second. This is done by repeatedly performing the \epc Inventory session (as enclosed in the `loop' box in \autoref{fig:ReaDmE_implementation}). $R_{\rm read}$ is, therefore, the total number of successful inventories over the time window.

\vspace{2mm}
\noindent\textbf{Stage 2.~}Calculate $T_{\rm c}$ using \autoref{eqn:tc_fn_of_RR}. To increase fault tolerance, we let the calculated sleep time $T_{\rm sleep} = \tau \times T_{\rm c}$, for example $\tau = 1.1$ gives a $10\%$ margin. In the sequence diagram, we denote this function as \textsf{ Get\_Tsleep($R_{\rm read}$)}.

\vspace{2mm}
\noindent\textbf{Stage 3.~}Calculated $T_{\rm sleep}$ is downloaded to the CRFID device via a standard \epc \texttt{Write} command. On the CRFID device, the received $T_{\rm sleep}$ is stored in a volatile register \textsf{SLEEPTIME}, and the IEM adjusts its LPM according to the received $T_{\rm sleep}$. 
We employed the open source code release from SecuCode~\cite{su2018secucode} at~\cite{secucodegithub} to integrate \readme within the IEM model described therein. In order to integrate \readme, two minor change needed to be made to the IEM implementation:
\begin{itemize}
    \item Server shall compute $T_{\rm sleep}$ based on the $R_{\rm read}$ and download it to the CRFID device, via a \texttt{Write} command to the \textsf{SLEEPTIME} register.
    \item The CRFID device shall handle the \texttt{Write} command addressing the new \textsf{SLEEPTIME} register and update the IEM settings according to the received data.
\end{itemize}

\section{Experimental validation}\label{sec:experiment}
We conducted the following three experiments to: i) validate our model regarding the relationship between the $R_{\rm read}$ and $T_{\rm c}$ (\autoref{sec:sensible}); ii) show that \readme is accurate (\autoref{sec:accurate}); and iii)  practical (\autoref{sec:practical}).

\begin{figure}[!ht]
    \centering
    \includegraphics[width=.8\linewidth]{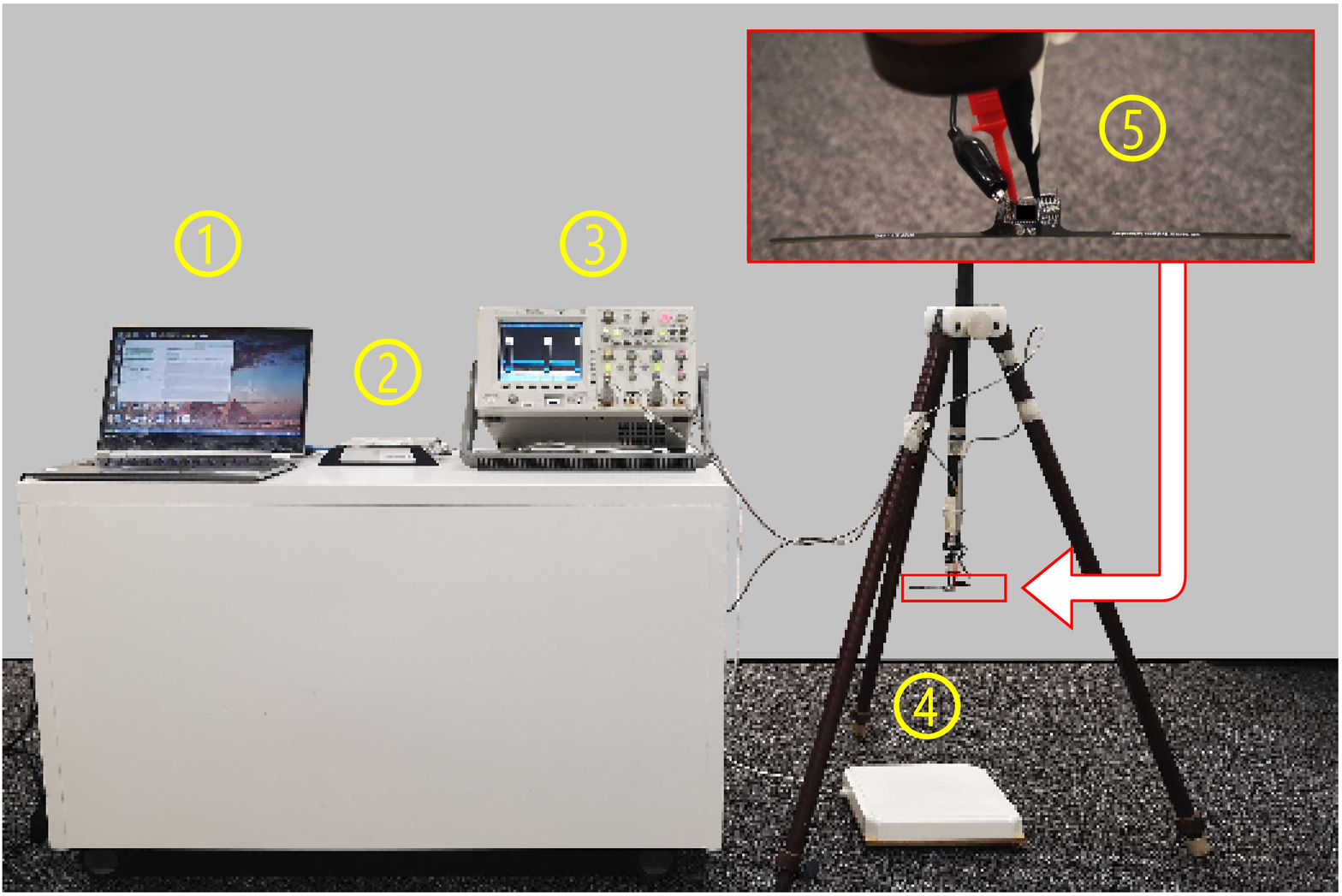}
    \caption{The experimental setup. \circled{1} a host PC ruining the Server Tool, \circled{2} an RFID reader (Impinj  R420  RFID), \circled{3} a digital storage oscilloscope (DSO), \circled{4} a reader antenna and \circled{5} a CRFID device suspended from a wooden tripod.}
    \label{fig:Exp_setup}
\end{figure}

All the experiments were conducted using the setup depicted in \autoref{fig:Exp_setup}. We had a laptop execute the Server tool, a modified version of the SecuCode Server tool at~\cite{secucodegithub}) that included the functions to count the read-rate and calculate the suitable $T_{\rm sleep}$ as described in \autoref{sec:design}. The laptop was connected to an Impinj R420 RFID reader, which drives a 9~dBic panel antenna. The CRFID sensor device being tested was suspended from a height adjustable wooden tripod. Instead of using the Joint Test Action Group (JTAG) debugging interface, we monitored a specific general-purpose input/output (GPIO) ping of the CRFID device with a digital storage oscilloscope (DSO) to measure the internal timing state. Because the JTAG provides power to the target device, it hinders us from investigating the CRFID device's behaviour under a passively powered condition. It also inserts extra debug-related code that interferes with our timing analysis. Hence, we programmed the CRFID device with a special firmware; when the CRFID device switches between its internal states---as shown in \autoref{fig:CRFID_Burst}~(b)---the firmware flips a specific GPIO pin corresponding to the current machine state.

\subsection{Validating our model}\label{sec:sensible}

The first experiment was conducted to collect and analyse all quantities mentioned in \autoref{sec:principle}. We sought to identify whether the $R_{\rm read}$ is a meaningful parameter reflecting CRFID device's behaviour, and further validate our theory and assumptions developed in \autoref{sec:principle}.

We recorded the read-rate $R_{\rm read}$(Reader) as the successful Inventory sessions per second from the reader side, averaged over 20 seconds. From the CRFID device side, we also measured timing information, such as charge time $T_{\rm c}$, active time $T_{\rm a}$ and transmit time $T_{\rm t}$. 

\begin{figure}[!ht]
    \centering
    \includegraphics[width=.7\linewidth]{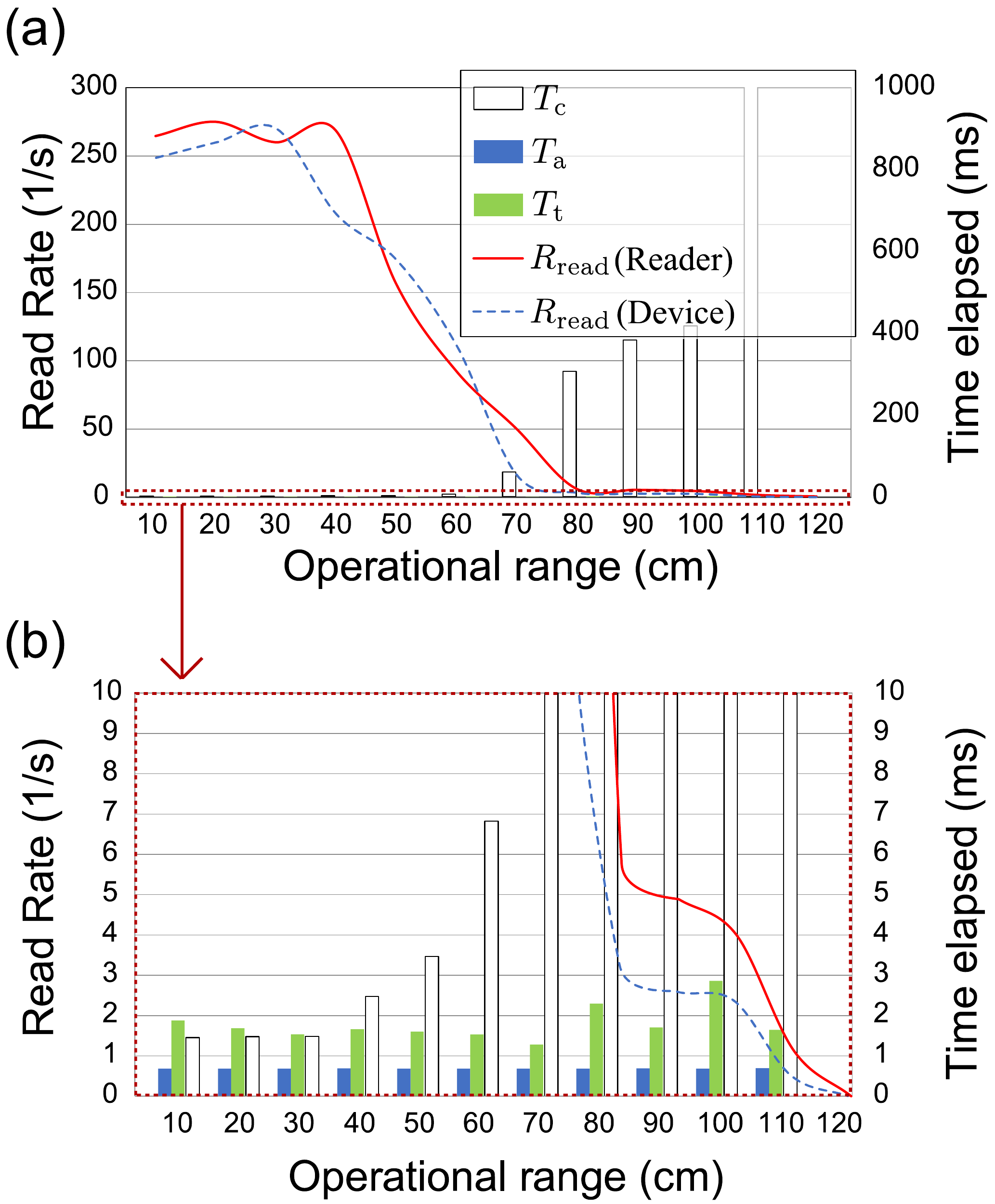}
    \caption{The time elapsed (right y-axis) for charging up $T_{\rm c}$, executing the program $T_{\rm a}$, transmitting data $T_{\rm t}$. On the left y-axis is the read-rate $R_{\rm read}$ reported from the RFID reader and calculated from on-device measurements using \autoref{eqn:RR2}). (a) Results for increasing operational range (x-axis). (b) A magnified view to better illustrate the measurements.}
    \label{fig:Tc_vs_Tc}
\end{figure}

The results are presented in \autoref{fig:Tc_vs_Tc}, with a white bar representing $T_{\rm c}$, a blue bar representing $T_{\rm a}$ and a green bar representing $T_{\rm t}$. We can see a trend representing an exponential increase in $T_{\rm c}$, while $T_{\rm a}$ and $T_{\rm t}$ remain approximately constant; in accordance with our assumptions described in \autoref{sec:thery_n_proof}. 

Using \autoref{eqn:RR2}, we can calculate the CRFID device's read-rate $R_{\rm read}$(Device), which is plotted with a blue dashed line in \autoref{fig:Tc_vs_Tc}. The $R_{\rm read}$(Device) is a good approximation of $R_{\rm read}$(Reader)---plotted by a red line in \autoref{fig:Tc_vs_Tc}. Notably, both the $R_{\rm read}$(Device) and $R_{\rm read}$(Reader) show a saturation for operation range between 10 and 30 cm. We can explain this observation by recognising that the harvested power at this close operating range is able to support the CRFID transponder to run continuously. Thus, the read-rate is dominated by $T_{\rm a}$ and $T_{\rm t}$. As the operational range increases, from 40 cm---see \autoref{fig:Tc_vs_Tc}.(b) for a clearer view---$T_{\rm c}$ follows an exponential trend. Meanwhile, $T_{\rm a}$ and $T_{\rm t}$ remain stable until the operational range reaches 80 cm. At this point, $T_{\rm t}$ becomes unstable, presumably due to power failures during the RFID backscatter under poor powering conditions.

The read-rate reported by the RFID reader closely follows the read-rate calculated by \autoref{eqn:RR2} based on $T_{\rm c}$, $T_{\rm a}$ and $T_{\rm t}$ measured directly on-device. We can confidently say that our model derived in \autoref{sec:thery_n_proof} is valid and the read-rate is a reliable reflection of the CRFID device's internal state.

\begin{table}[!ht]
  \centering
  \caption{Measured values for Equation~\eqref{eqn:tc_fn_of_RR}.}\label{tab:exp_constante}
  \resizebox{.8\linewidth}{!}{
      \begin{tabular}{cccc}
        \toprule
         & mean ($\mu$) & standard deviation ($\sigma$)  \\\hline\hline
        CPU active time ($T_{\rm a}$) & 0.630 & 0.190  \\
        Data transmission time ($T_{\rm t}$) & 1.641 & 0.635 \\
        $K = T_{\rm a} + T_{\rm t}$ & 2.271 & 0.792 \\
        \bottomrule
      \end{tabular}
   }
\end{table}

From this experiment, we can also find the constant $K$ in \autoref{eqn:tc_fn_of_RR}, as $K = T_{\rm a} + T_{\rm t}$. The average over all tested operational ranges from \autoref{fig:ReadRateChargeTime} are summarised in \autoref{tab:exp_constante}. This reaffirms our assumption in \autoref{sec:principle} that $T_{\rm a}$ and $T_{\rm t}$ are invariant over different powering conditions.

\subsection{The accuracy of read-rate based charging time predictions}\label{sec:accurate}
\begin{figure}[!ht]
    \centering
    \includegraphics[width=.7\linewidth]{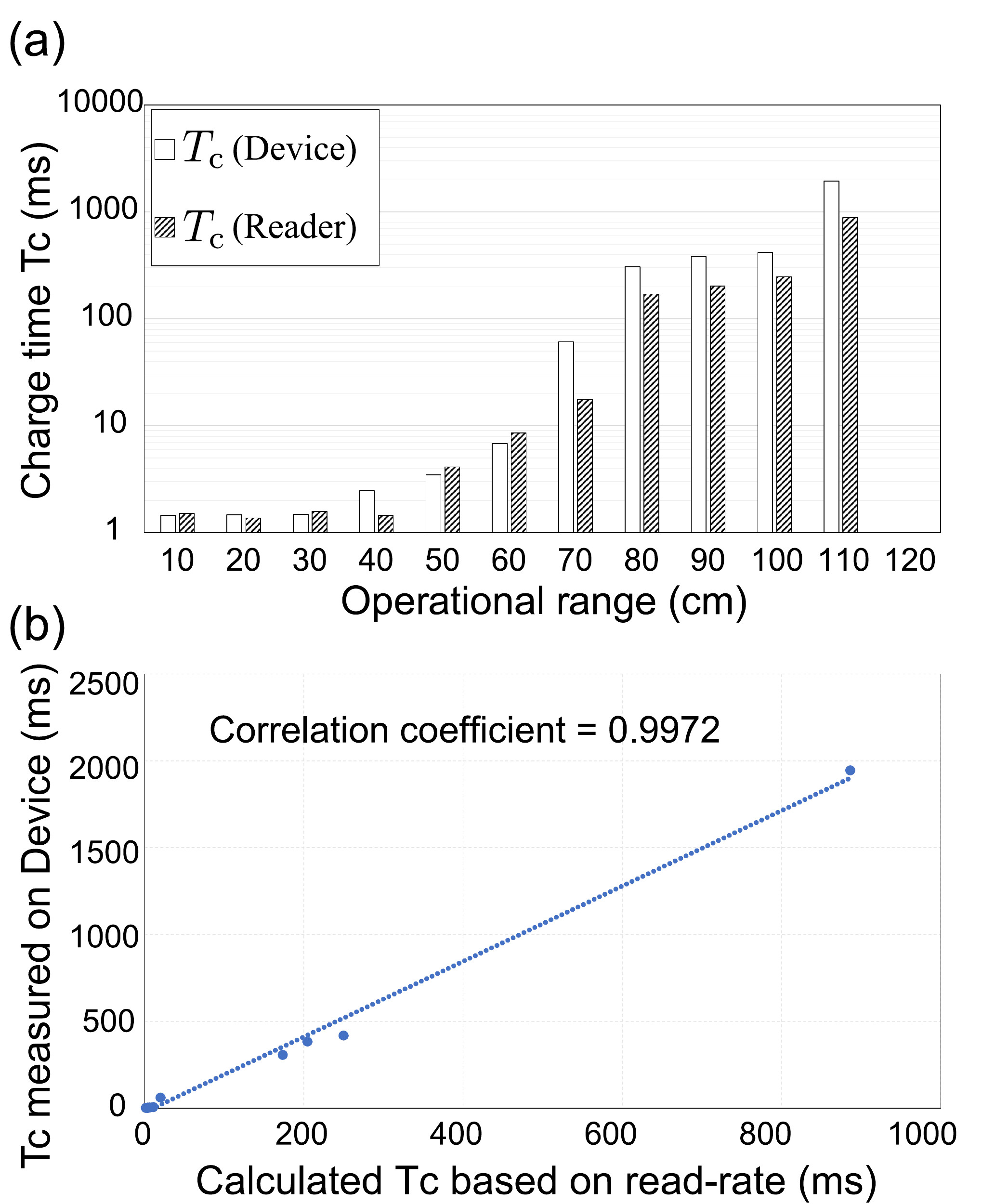}
    \caption{(a) Comparing the actual charging time measured from the device $Tc$~(Device) and the charging time calculated based on the $R_{\rm read}$ $Tc$~(Reader) with \autoref{eqn:tc_fn_of_RR}, and (b) the correlation between the $Tc$~(Device) and the $Tc$~(Reader) evaluations.}
    \label{fig:ReadRateChargeTime}
\end{figure}
From \autoref{fig:ReadRateChargeTime}, we can see that the $T_{\rm c}$ measured directly from the device closely follows that evaluated using \autoref{eqn:tc_fn_of_RR} using the $R_{\rm read}$ reported by the RFID reader. However, a systematic analysis is still required to determine whether $R_{\rm read}$ is an accurate measurement of $T_{\rm c}$.

To examine the accuracy of the read-rate based powering condition measurement, we apply the Pearson correlation test \cite{schober2018correlation}. We use the $T_{\rm c}$ measured from the device as one tested variable and use the $T_{\rm c}$ calculated based on $R_{\rm read}$ as the other tested variable. The result is plotted in \autoref{fig:ReadRateChargeTime}~(b). We can see that all measured points are located very close to the trend line. The Pearson correlation coefficient of 0.9972 clearly shows that there is a solid correlation between the $T_{\rm c}$ directly measured from the device and the $T_{\rm c}$ calculated based on $R_{\rm read}$ from the RFID reader using \autoref{eqn:tc_fn_of_RR}.

\subsection{Application case study with a cryptographic algorithm}\label{sec:practical}
To demonstrate the effectiveness of our \readme method, we followed the same setup as in~\cite{su2018secucode} and performed a message authentication code (MAC) over a 1280-byte random string, and compared against CEM and IEM with hard-coded 30~ms $T_{\rm sleep}$. We repeated each measurement ten times and report the success rate (successfully calculate the MAC tag) and latency (average time taken). Notably, data was only collected when MAC computation commenced, where the DSO was triggered by a GPIO event. Failures in underlying RFID communications were ignored---notably, \readme cannot be applied to RFID communications, as the injected LPM can violate the strict RFID communication timing requirements. 

\begin{figure}[!ht]
    \centering
    \includegraphics[width=.7\linewidth]{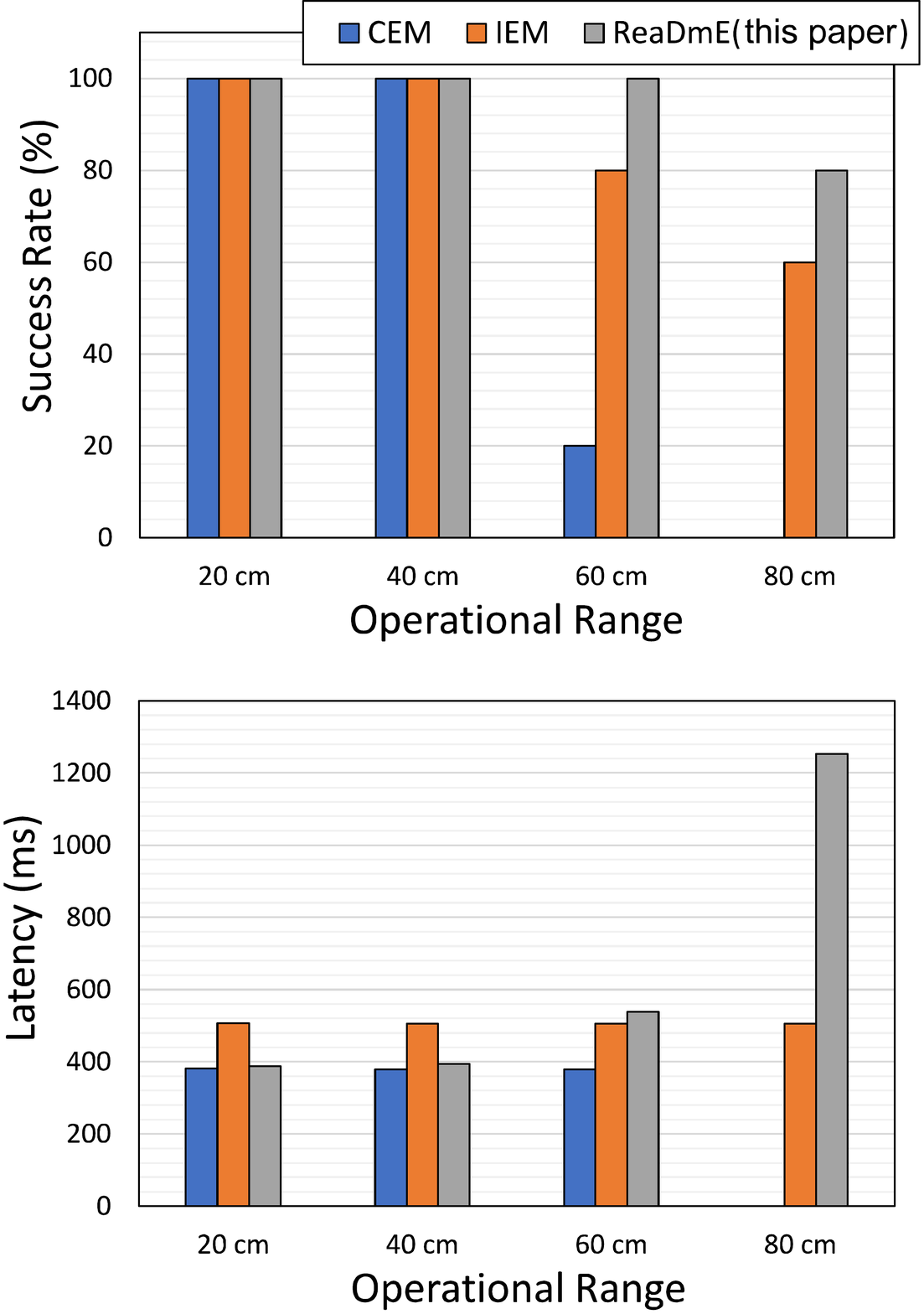}
    \caption{Experiment results comparing the success rate and the latency across the continuous operational model (CEM), the intermittent operational model (IEM), and our proposed ReaDmE.}
    \label{fig:Exp_result}
\end{figure}

The experimental results are summarised in \autoref{fig:Exp_result}. On the one hand, in terms of latency, \readme demonstrated  very close performance to the CEM when power is plentiful at 20~cm and 40~cm, while IEM suffered from large unnecessary time overheads due to the hardcoded 30~ms dead weight. On the other hand, \readme achieved improved success rate due to the dynamically generated LPM duration based on an accurate assessment of the power available to the CRFID device. It is worth pointing out that \readme could achieve a 100\% success rate at a 60 cm operational range while IEM fails on 20\% of its trials at this range. \readme was also 20\% more successful than the IEM under the poor powering condition at an 80~cm operational range. Notably, as shown in Fig. 8, the CEM’s success rate dropped to $20\%$ at a 60~cm range and to $0\%$ at 80~cm range.

\section{Conclusion and Future work}\label{sec:conclusion}
We have shown that the read-rate, reported from the host reader, could be used as an accurate estimate of powering channel condition at an RF-powered device. We presented our \readme scheme for estimating the required charging time for a CRFID  device under intermittent powering conditions. We applied \readme to build a dynamic execution scheduling method for a CRFID type passively powered device. Empirical results showed that \readme can handle the power-loss more efficiently. Under good powering conditions, \readme reduced timing overhead by $23\%$ and \readme was 20\% more successful than its competitor, IEM, under critical powering conditions. More importantly, \readme requires no hardware modification to the standard CRFID architecture whilst placing minimal overheads on the device.

Although we have shown that the read-rate is an excellent measure in a single-device setup and under reduced multi-path conditions, the channel is likely to be more complicated in situations with more devices, considering potential device-to-device interference, collisions,  throughput limitation of an ALOHA channel\cite{xu2015improved}, as well as those channels heavily affected by multi-path fading. In the future, we will further investigate the read-rate based method in a multi-device setup.

Other than the read-rate, there are numerous features we can extract from the RFID communication channel, such as RSSI, CRC (cyclic redundancy check) error rate, device collision rate and the success rate of writing to device memory. We will consider such additional data sources in our future work.

\section{Acknowledgement}
This work was partially supported by the Australian Research Council Discovery Program (DP140103448). We also want to express our gratitude to Mr. Michael Chesser, for his feedback and suggestions as well as making the necessary modification to the Server Tool to conduct the experiments.

\bibliographystyle{IEEEtran}
\bibliography{main}

\end{document}